\documentclass[aps,prb,showpacs,twocolumn,longbibliography]{revtex4-1}

\usepackage[english]{babel}
\usepackage[utf8]{inputenc}
\usepackage{amsmath}
\usepackage{amssymb}
\usepackage[caption=false]{subfig}
\usepackage{amssymb}
\usepackage{epsfig}
\usepackage{graphicx}
\usepackage{amsmath}
\usepackage{array,color}
\usepackage{natbib}

\usepackage[usenames,dvipsnames]{xcolor}
\definecolor{forestgreen}{rgb}{0.11,0.54,0.15}
\definecolor{purple}{rgb}{0.62,0.10,0.96}
\definecolor{dockerblue}{rgb}{0.11,0.56,0.98}
\definecolor{freeblue}{rgb}{0.25,0.41,0.88}

\usepackage[pdftex,plainpages=false,colorlinks=true,linkcolor=Red, citecolor=blue, urlcolor=blue]{hyperref}

\begin{document}
\title{Orbital Effect of the Magnetic Field in Dynamical Mean-Field Theory}
\author{S. Acheche$^{1*}$, L-F. Arsenault$^{1*}$ and A.-M. S. Tremblay$^{1,2}$}
\affiliation{
$^1$D\'{e}partement de physique, Institut quantique, and Regroupement qu\'eb\'ecois sur les mat\'eriaux de pointe, Universit\'{e} de Sherbrooke, Sherbrooke, Qu\'{e}bec, Canada J1K 2R1 \\
$^2$Canadian Institute for Advanced Research, Toronto, Ontario, Canada, M5G 1Z8\\
$*$ These authors contributed equally on this work
}
\date{\today}
\begin{abstract}
The availability of large magnetic fields at international facilities and of simulated magnetic fields that can reach the flux-quantum-per-unit-area level in cold atoms, calls for systematic studies of orbital effects of the magnetic field on the self-energy of interacting systems. Here we demonstrate theoretically that orbital effects of magnetic fields can be treated within single-site dynamical mean-field theory with a translationally invariant quantum impurity problem. As an example, we study the one-band Hubbard model on the square lattice using iterated perturbation theory as an impurity solver. We recover the expected quantum oscillations in the scattering rate and we show that the magnetic fields allow the interaction-induced effective mass to be measured through the single-particle density of states accessible in tunneling experiments. The orbital effect of magnetic fields on scattering becomes particularly important in the Hofstadter butterfly regime. 
\end{abstract}
%\pacs{71.27.+a, 72.10.-d, 72.10.Bg, 72.15.Jf, 72.20.Pa}
%  71.27.+a 	Strongly correlated electron systems; heavy fermions
% 72.10.-d	Theory of electronic transport; scattering mechanisms
% 72.10.Bg	General formulation of transport theory 
% 72.15.Jf	Thermoelectric and thermomagnetic effects (metals and alloys)
% 72.20.Pa	Thermoelectric and thermomagnetic effects (semiconductors and insulators)
\maketitle

%----------------------------------------------------------------------------------------
%	ARTICLE CONTENTS
%----------------------------------------------------------------------------------------

%\begin{multicols}{2} % Two-column layout throughout the main article text

\section{Introduction} 

Dynamical Mean-Field Theory (DMFT)\cite{Georges_1992, Georges_1996} is one of the most successful methods that deals with electron-electron correlations. Consistency between local atomic multiplets and extended lattice states is the main conceptual idea behing this theory. This is achieved by solving a quantum impurity problem whose hybridization function is determined self-consistently through the requirement that the Green function of the impurity is the one that can be obtained by projecting on the impurity the lattice Green function with the same frequency-dependent self-energy as the impurity. Exact at infinite dimension, its major achievement was to accurately describe the so-called Mott transition \emph{i.e.} a metal-to-insulator transition due to interactions. This theory is also used to describe broken-symmetry phases, such as antiferromagnetism, ferromagnetism or superconductivity, making DMFT a relevant choice to study three-dimensional correlated materials. 
%To deal with low-dimensional physics where spatial quantum correlations must be well described, cluster extensions based on the same philosophy as DMFT were developed. These extensions include cluster dynamical mean-field theory (CDMFT)\cite{Kotliar_2001, Lichtenstein_2000} and the dynamical cluster approximation (DCA)\cite{Maeir_2005}.\\

The orbital effect of magnetic fields on electrons moving on a lattice is non trivial, even in the absence of interactions. Beyond the semiclassical picture of cyclotronic closed orbitals that occur at low magnetic fields in parabolic bands, the presence of a periodic potential can completely modify the energy of Bloch electrons. In two dimension (2D), the most spectacular effect of this modification is the appearance of the famous Hofstadter butterfly\cite{Hofstadter_1976, MacDonald_1983} when energy levels are calculated as a function of the magnetic flux per plaquette in units of the magnetic flux quantum $\Phi_0 = h/e$. The appearance of this fractal structure is most apparent when this dimensioneless magnetic flux is a rational number $p/q$ with $q$ not too large. This structure is directly linked to the presence of competing lattice and magnetic-flux-per-plaquette periodicity. 
Unfortunately, this physics is only accessible at  unattainable magnetic fields (of the order of $10^5$ T) in real materials. However, artificial structures, such as cold atom lattices or Moir\'e superlattice, allow one to experimentaly realize the Hofstadter Butterfly\cite{Goldman_2010,Gerbier_2010,Dean_2013}.

The effect of interactions on  Hofstadter's butterfly has already been studied by various methods such as mean-field theory \cite{Gudmundsson_1995,Mishra_2016, Doh_1998}, DMFT for the Falicov-Kimball model\cite{Tran_2010} or real-space DMFT\cite{Cocks_2012, Kumar_2016, Orth_2013}. The latter approach generalizes DMFT\cite{Potthoff_1999, Snoek_2008} to the case where the electromagnetic vector potential breaks translational invariance, by using a set of quantum impurities, one for each inequivalent site of the lattice. 

Although DMFT has been used to describe the effect of a uniform magnetic field with a single-impurity problem in the Falicov-Kimball model,\cite{Tran_2010} the general form of the DMFT self-consistency equation in a magnetic field has not been proven. In Sec.\ref{Sec:DMFT}, we derive the DMFT equations in the case where a uniform magnetic field is applied. This derivation is gauge independent and works in any dimension and for any lattice geometry. In Sec.\ref{Sec:MethodModel}, we introduce the impurity solver that we use in this work for the DMFT calculation. The results for the square lattice Hubbard model with nearest-neighbor hopping in the presence of a magnetic field, in Sec.\ref{Sec:Results}, show that at low magnetic field, far from half-filling, one recovers Lifschitz-Kozevitch theory, \emph{i.e.} Landau levels and quantum oscillations in measurable quantities. These oscillations are observable as well in the electron lifetime, which is so important for Shubnikov De Haas oscillations. At half-filling, which corresponds to a Fermi energy that lies on a Van-Hove singularity, the signature of Hofstadter physics are clearly visible in the electron lifetime, which would influence Shubnikov de Hass oscillations for example. Our work is particularly relevant for transport measurements of cold atoms in optical lattices.

\section{The Dynamical Mean-Field equations with magnetic field}\label{Sec:DMFT}
Here we derive~\cite{Arsenault:these} the DMFT equations when the orbital effect of the magnetic field is taken into account in the Hubbard model.
%\footnote{This derivation is based on the PhD thesis of Louis-Fran\,cois Arsenault, Universit\'e de Sherbrooke, 2013} 
The effect of a Zeeman coupling \cite{LalouxGeorges:1994} is trivial to introduce and will not be discussed here. As usual, the orbital effect of the applied magnetic flux is taken into account through the Peierls substitution which, in second quantization, consists in a change of the hopping term in the kinetic part of the Hamiltonian. The Hubbard model takes the form:
\begin{equation}
H = - \sum_{m,n, \sigma} t_{mn} e^{if_{mn}} \hat{c}^\dagger_{m \sigma} \hat{c}_{n\sigma} + U \sum_{m} \hat{n}_{m\uparrow} \hat{n}_{m\downarrow} - \mu \sum_{m\sigma} \hat{n}_{m\sigma}.\label{Eq:Hamiltonian}
\end{equation}
Here, $\hat{c}_{m\sigma}$ ($\hat{c}^\dagger_{m\sigma}$) destroys (creates) an electron of spin $\sigma$ on site $m$ and $\hat{n}_{m\sigma}$ is the number operator for electrons of spin $\sigma$ on site $m$, while $U$ and $\mu$ are respectively the on-site Coulomb repulsion and the chemical potential. The hopping amplitude between sites $m$ and $n$ is described by the real-symmetric matrix $t_{mn}$. The magnetic flux is taken into account through the Peierls phase $f_{mn}$:
\begin{equation}
f_{mn} = \frac{2\pi}{\Phi_0}\int^n_m  \textbf{A}( \textbf{r})\cdot d\textbf{l}
\end{equation} 
where $\textbf{A}(\textbf{r})$ is a vector potantial corresponding to the external magnetic flux density $\textbf{B}$ and $\Phi_0=h/e$ the magnetic flux quantum. The line integral is along a straight line connecting sites $m$ and $n$. We adopt natural units where \emph{i.e} $\hbar=k_B=1$. Although in the next section we present numerical results for the two-dimensional square lattice, the derivation in this section is valid for an arbitrary lattice in arbitrary dimension. 

In the following, we denote the generalized hopping term $t_{mn}\exp(i f_{mn})$ as $\tilde{t}_{mn}$ to simplify the notation and derive the DMFT equations. The lattice self-energy can be written in the following form: $\Sigma_{mn}(i\omega_n, \textbf{B})= e^{if_{mn}}\Bar{\Sigma}_{mn}(i\omega_n, \textbf{B})$ where $\Bar{\Sigma}$ is a translation and gauge invariant self-energy\cite{Takafumi:2005}. Rewritting the self-energy in this form proves us that, even if we take the local part of the self-energy, \emph{i.e} $m=n$, dependancy on the external magnetic field is still present. This statement is particulary important in DMFT since it approximates the lattice self-energy by a purely local impurity self-energy. Since the local self-energy is in principle measurable through lifetimes or densities of states, it cannot depend on position because the magnetic field and the lattice are uniform. Although it is not a rigorous proof, a fourth-order perturbative developpement of the self-energy shows that the on-site self-energy does not depend on the site.~\cite{Note1} 

Knowing that, Dyson's equation relating the interacting and the non-interacting Green's functions can be rewritten in a useful way. Setting $U=0$ in the Hamiltonian Eq.~\ref{Eq:Hamiltonian}, and using matrix notation in the space of site indices, the equation of motion for the non-interacting Green function $G^0_{mn}$ is,
\begin{eqnarray}
(i\omega_n+\mu)\mathbf{I}\textbf{G}^{0} &=& \mathbf{I} - \tilde{\mathbf{t}}\mathbf{G}^0 \\
\mathbf{G}^0(i\omega_n) &=& \left[(i\omega_n+\mu)\mathbf{I} + \tilde{\mathbf{t}} \right]^{-1} \label{Eq.G0}
\end{eqnarray}
where $\mathbf{G}^0$ is the matrix whose the $m,n$ element is $G^0_{mn}$, $\tilde{\mathbf{t}}$ is the hopping matrix in presence of magnetic field and $\mathbf{I}$ the identity matrix. Using Dyson's equation in the case of a local and site-independant self-energy, we find
\begin{eqnarray}
\mathbf{G}^{\text{int}} &=& \mathbf{G}^0 + \mathbf{G}^0 \left[\Sigma(i\omega_n, \textbf{B}) \mathbf{I}\right] \mathbf{G}^{\text{int}} \\
&=&  \left[(\mathbf{G}^0)^{-1} - \Sigma(i\omega_n,\textbf{B})\mathbf{I} \right]^{-1}\\
&=& \left[(i\omega_n+\mu - \Sigma(i\omega_n,\textbf{B}))\mathbf{I} +\tilde{\mathbf{t}} \right]^{-1} \\
&=& \mathbf{G}^0(i\omega_n -\Sigma(i\omega_n,\textbf{B}))
\end{eqnarray}
where we used Eq~\eqref{Eq.G0}. This link between the non-interacting and interacting Green's functions is very convenient since  the equation of motion of the latter takes the simple form,

\begin{equation}
(i\omega_n+\mu-\Sigma(i\omega_n))G_{mn}(i\omega_n) =\delta_{mn}-\sum_k t_{mk} e^{if_{mk}}G_{kn}(i\omega_n).\label{EqMvt}
\end{equation} 

To derive the DMFT equations, we use the cavity method.\cite{Georges_1996} The basic idea of this method is to divide the lattice problem in two parts: The lattice in the presence of a cavity (\emph{i.e.} the absence of one site) and the cavity site. After integrating out the degrees of freedom of the lattice in the presence of the cavity, one can obtain the dynamics for the cavity site.

The partition function of the Hubbard model with magnetic field can be written as a functionnal integral over Grassmann variables,
\begin{equation}
Z = \int \Pi_{m,\sigma} {\cal{D}} c^\dagger_{m\sigma} {\cal{D}}c_{m\sigma} e^{-S}\, .
\end{equation}
At finite temperature, the action $S$ can be written as an integral over imaginary time $\tau$:
\begin{eqnarray}
S &=&\int^{\beta}_0 d \tau \left[\sum_{m,\sigma} c^\dagger_{m\sigma}(\tau)\left(\frac{\partial}{\partial \tau}-\mu \right)c_{m \sigma}(\tau)\right. \\
\nonumber
&-& \left. \sum_{m,n,\sigma} \tilde{t}_{mn}c^\dagger_{m \sigma}(\tau)c_{n \sigma}(\tau)+U \sum_{m} n_{m \uparrow}(\tau) n_{m\downarrow}(\tau)\right].
\end{eqnarray}
By construction, the cavity method divides the action into three parts: the action of the lattice with the cavity, the action of the cavity site, which is from now on denoted with the $l$ index, and the action of the hybridization between the cavity and the lattice. The latter piece of the action takes the form

\begin{eqnarray}
\nonumber
\Delta S = &-& \int^\beta_0 d\tau \sum_{\sigma,i}\left[ \tilde{t}_{il}c^\dagger_{i\sigma}c_{l\sigma}+\tilde{t}_{li}c^\dagger_{l\sigma}c_{i\sigma}\right] ,\\
\end{eqnarray}
where the sum over $i$ does not include the cavity.

Following the steps of the usual derivation of the dynamical mean-field equations for the cavity method we use the linked cluster theorem and obtain the effective action at the site of the cavity,
\begin{eqnarray}
\nonumber
S_{eff,l} = &-& \int^\beta_0 d\tau_1 \int^\beta_0 d\tau_2\sum_{\sigma} c^\dagger_{l\sigma}(\tau_1)\mathcal{G}^{-1}_{0,l}(\tau_1 -\tau_2)c_{l\sigma}(\tau_2)\\
&+& U\int^\beta_0 d\tau n_{l\uparrow}(\tau) n_{l\downarrow}(\tau). \\
\nonumber
\end{eqnarray}
Here, $\mathcal{G}^{-1}_{0,l}(\tau_1-\tau_2)$ plays the role of a Weiss effective field at the cavity site whose expression in terms of Matsubara frequencies and interacting lattice Green function $G^{l}$ with the $l$ site missing, is, 
\begin{eqnarray}\label{Eq.Weiss}
\mathcal{G}^{-1}_{0,l}(i\omega_n) &=&i\omega_n +\mu -\frac{1}{2}\sum_{jk} \left( \tilde{t}_{lj} G^{l}_{jk}(i\omega_n)\tilde{t}_{kl}\right.\\
\nonumber
&+& \left. \tilde{t}_{lk}G^{l}_{kj}(i\omega_n)\tilde{t}_{jl} \right)\\
\nonumber
&=& i\omega_n +\mu -\sum_{jk}  \tilde{t}_{lj} G^{l}_{jk}(i\omega_n)\tilde{t}_{kl}.
\end{eqnarray}
We dropped out the spin index $\sigma$ for the sake of clarity. This expression is similar to the one presented in Ref.\onlinecite{Georges_1996}. The last term of Eq.\eqref{Eq.Weiss} can be seen as an hybridization function between the cavity and the lattice without the cavity site, making the impurity solvers for DMFT available in presence of magnetic field.

The next step is to express the effective Weiss field in terms of $G$, the lattice Green function with all the sites present, and then to write the resulting expression in terms of the cavity-site Green function in order to find the self-consistency relation. In Eq.~\eqref{Eq.Weiss}, we use the formula that links the lattice Green's function in the presence of the cavity with the lattice Green's function without the cavity~\cite{Hubbard:1964}
\begin{equation}
G^l_{i j}=G_{ij}-\frac{G_{il}G_{lj}}{G_{ll}}
\label{Eq.Green_l}
\end{equation}
and make repeated use of the equation of motion Eq.~\eqref{EqMvt} to finally obtain the Weiss field in the following form:
\begin{equation}
\mathcal{G}^{-1}_{0,l} (i\omega_n) = i\omega_n +\mu + \sum_n \frac{\tilde{t}_{ln}G_{nl}}{G_{ll}}.\label{EqWeiss}
\end{equation}
It is clear that in the presence of magnetic field, the Green's function is not translationnally invariant, making the Weiss field not trivial to simplify at first sight. Yet, one can construct a translation and gauge invariant Green's function $\bar{G}$ by using the following transformation $\bar{G}_{mn} = e^{-if_{mn}}G_{mn}$\cite{Chen_2011, Khodas_2003}. Using this result, the equation of motion for the cavity site Green function Eq.\eqref{EqMvt} gives us the relation between $G_{ll}$ and $\bar{G}_{ln}$ 
\begin{equation}
(i\omega_n +\mu - \Sigma(i\omega_n))G_{ll}(i\omega_n) = 1 - \sum_{n} t_{ln} \bar{G}_{nl}(i\omega_n).
\end{equation}
Substituting the right-hand side of this equation in the equation for the Weiss field Eq.~\eqref{Eq.Weiss} gives the following translationally invariant final form for the Weiss field 
\begin{equation} \label{{Eq.Weiss_final_dmft}}
\mathcal{G}^{-1}_0(i\omega_n) = \Sigma(i\omega_n) + (G_{ll}(i\omega_n))^{-1}.
\end{equation}
This expression is convenient because it is exactly the same expression as in the absence of magnetic field. The effect of the magnetic field on the self-energy and the hybridization function only comes from the non-interacting density of states, allowing the use of all existing impurity solvers. To our knowledge, this statement was often affirmed without rigorous proof. 

In closing, note that once the hypothesis of a site-independent local self-energy is accepted, the self-consistency equation also follows simply within a Luttinger-Ward formalism. Indeed, the local Green function entering the calculation of the Luttinger-Ward functional is the same as the local Green function obtained from the projection of the lattice Green's function that contains that self-energy. 

\section{method and model}\label{Sec:MethodModel}

As pointed out above, the effect of the magnetic field is contained in the local Green's function $G_{ll}$, therefore, the strategy to solve the DMFT problem is to compute the eigenvalues of the non-interacting lattice in the presence of uniform magnetic field and use them in the self-consistency loop. There are numerous methods to solve the non-interacting case\cite{Hofstadter_1976, Berciu_2010, Ueta_1997}. For the numerical example presented below, we take the case of a uniform magnetic field perpendicular to a square lattice and we compute the energy level of Bloch electrons by solving an almost Mathieu equation named Harper equation \cite{Harper_1955} for rational ratios of magnetic flux, \emph{i.e.} $eBa^2/h = p/q$ where $a$ is the lattice constant (taken as unity) and $p$ and $q$ are coprime integers. This choice looks arbitrary at first sight but it allows to define a commensurable magnetic unit cell and simplifies at the same time the computational work. The question of rational or irrational ratio has been discussed since the original paper of Hofstadter\cite{Hofstadter_1976}. It is important to stress that our derivation of the DMFT equation does not depend on this choice. In practice, in the Landau gauge and in a model with only nearest-neigbor hopping $t$, we obtain the following Harper equation for the wave function on the sites of the magnetic unit cell
\begin{eqnarray}\label{eigPb}
\psi_{n+1}+\psi_{n-1} +2\cos (2\pi n \frac{p}{q}-k_y)\psi_n = \frac{\epsilon}{t} \psi_n.\label{Eq:Harper}
\end{eqnarray}
Here $n=1\ldots q$ indexes the sites in the magnetic cell and $k_y$ is the momentum along the $y$ axis. For realistic magnetic fields, say to the order of one Tesla, one must diagonalize a $10^5 \times 10^5$ matrix, which is out of reach for numerical methods. This constraint can be bypassed by using perturbation theory, but we will rather focus on values of q corresponding to an intermediate regime between Landau levels physics and Hofstadter butterfly physics, \emph{i.e.} q of the order of hundreds. This regime has two advantages: The eigenproblem Eq.\eqref{eigPb} is then easily solvable numerically and the energy between two eigenvalues is large enough compared to the lowest temperature we can reach that effects of the magnetic field are not washed out by thermal effects.

In order to solve the Anderson impurity problem, we use Iterated Perturbative Theory (IPT)\cite{Kajueter_1996,Zhang_1993}. We have checked the consistency of our results with a Continous Time Quantum Monte-Carlo solver\cite{Haule_2007, Gull_2011}. IPT is an interpolation method between second order perturbation theory and the atomic limit for the impurity self-energy. Although it was one of the first impurity solvers used in DMFT, IPT captures qualitatively the main physics of the Hubbard model. However, it has issues when the system studied is far from half-filling and in the strong interaction limit.~\cite{Arsenault:2012} The IPT solver has the advantage of being more easily analytically continued in the present context where effects are often small.

The IPT self-energy for a given spin $\sigma$ at a finite temperature $T$ has the following expression in Matsubara frequencies $i\omega_n$: 
\begin{equation}
\Sigma_\sigma (i \omega_n) = U\frac{n}{2} +\frac{A\Sigma^{(2)}_\sigma (i\omega_n)}{1-B\Sigma^{(2)}_\sigma (i\omega_n)}
\end{equation}
with
\begin{equation}
\Sigma^{(2)}(i \omega_n)_\sigma = -U^2 \int^\beta_0 \mathcal{G}^\sigma_0(\tau) \mathcal{G}^{-\sigma}_0 (-\tau)\mathcal{G}^{-\sigma}_0 (\tau)d\tau 
\end{equation}
where
\begin{equation}
\mathcal{G}_0(i \omega_n) = \frac{1}{i\omega_n +\mu_0 -\Delta (i\omega_n)}.
\label{Eq:Weiss} 
\end{equation}

Note that, in the absence of a Zeeman term, the only link between the self-energy and the magnetic field lies in the hybridization function $\Delta (i\omega_n)$. From a physical point of view, the Green's function $\mathcal{G}_0$ corresponds to the amplitude for a particle to return to the impurity after a voyage in the bath. Finally, the constants $A$ and $B$ are chosen in such a way that the self-energy in the strong coupling regime far from half-filling is exact in the atomic limit and has the correct high-frequency behavior, that is
\begin{eqnarray}
A &=& \frac{n(2-n)}{n_0(2-n_0)} \\
B &=& \frac{\left(1- \frac{n}{2}\right)U+\mu_0 -\mu}{\frac{n_0}{2}\left(1-\frac{n_0}{2}\right)U^2}.
\end{eqnarray}
Here,  $n = G(\tau=0^-)$ and $n_0 = G_0(\tau=0^-)$ while $\mu$ and $\mu_0$ are the corresponding chemical potentials for the densities: $n_0$ has no physical meaning far from half-filling but it is taken equal to $n$. The latter corresponds to the electronic density of the lattice, the lattice Green's function being
\begin{equation}
G (i \omega_n) = \sum_m \frac{1}{i\omega_n + \mu -\epsilon_m - \Sigma(i\omega_n)}
\end{equation}
with $\epsilon_m$ the single-particle excitation energies of the non-interacting system.\\

The IPT implementation requires the use of Fast Fourier Transforms. Thus, we need to increase the convergence of the sums over Matsubara frequencies of Green functions. One way to achieve it consists in substracting and adding the asymptotic high-frequency behavior of the Green function. The high-frequency behavior of the hybridization function up to order $(i\omega_n)^{-1}$ and that of the local Green runction up to order $(i\omega_n)^{-3}$ are independent of magnetic field, which simplifies the calculation. This independence can be proven~\cite{Arsenault:these} by an easy generalization of the procedure in Ref.~\onlinecite{Koch:2008}. This is discussed in Appendix~\ref{Appendix:asymptotics}. 

\section{results}\label{Sec:Results}
We first present the results for the self-energy in Matsubara frequencies, then we show that oscillations in the local density of states allow one to extract the effective mass. We end with results for the field dependence of scattering time in the normal state.

\subsection{Matsubara frequency results}

\begin{figure}[h!]
\centering
\includegraphics[width=0.48\textwidth]{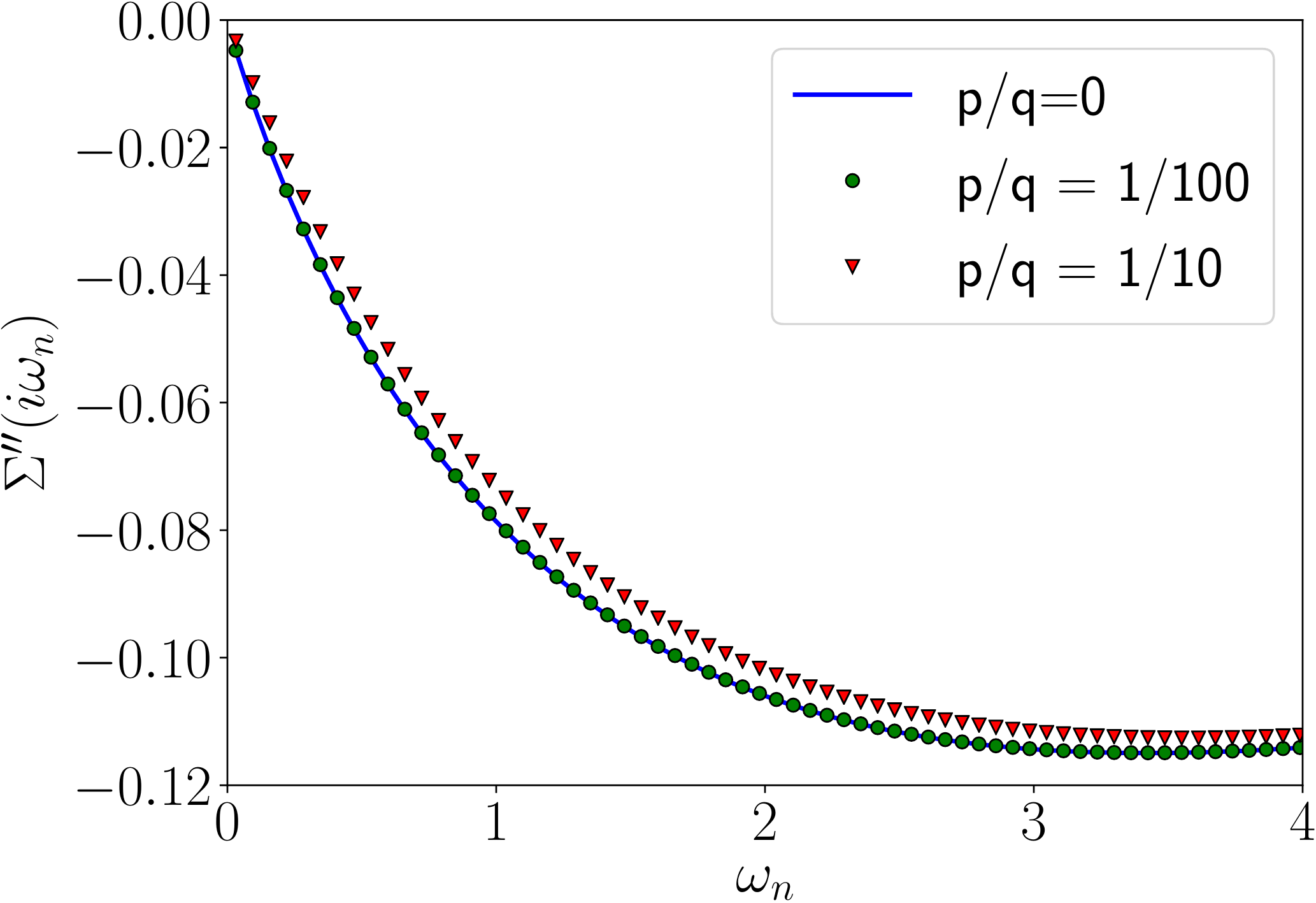}
\caption{(color online) Local self-energy as a function of Matsubara frequencies for $\beta=100$ and $U=2$ at half-filling in units where hopping $t$ is unity. The continous line interpolates the solution without magnetic field. Green circles ($1/q=1/100$) and red diamonds ($1/q=1/10$) represent the self-energy for a magnetic flux per plaquette in units of the flux quantum $eBa^2/h=\Phi/\Phi_0=1/q$.}\label{Fig.Self_iwn}
\end{figure}

We take the hopping term $t$ as unit of energy. In Fig.\ref{Fig.Self_iwn}, we plot the imaginary part of the local Matsubara self-energy for different dimensionless magnetic fluxes at $\beta=100$, $U=2$ and half-filling, $n=1$. When the magnetic flux per plaquette in units of the flux quantum approaches unity, $\Phi/\Phi_0=1/10$, the difference with the $\Phi/\Phi_0=0$ case is clear. This is Hofstadter's regime \emph{i.e.} the regime where, due to the $k_y$ dependency of eigenvalues of the Harper matrix, the density of states has drastic changes of topology, leading to a rich physics which is well captured by DMFT and which is visible for this value of $\beta$. Although we give more results in this regime when we discuss scattering-rate below, we do not enter into the many details for this case since numerous studies have already tackled the interacting Hofstadter butterfly.\cite{Gudmundsson_1995,Mishra_2016, Doh_1998, Tran_2010,Cocks_2012, Kumar_2016, Orth_2013}

In the Landau regime, \emph{i.e.} when the $k_y$ dependence of the eigenvalues in the Harper matrix Eq.\eqref{Eq:Harper} can be neglected, the self-energy appears unmodified compared to the $\Phi/\Phi_0=0$ case, as can be seen for $\Phi/\Phi_0=1/100$ in Fig.\ref{Fig.Self_iwn}: The self-energy has only minor corrections compared with the $\Phi/\Phi_0=0$ case. These minor corrections can be tracked by changing the magnetic field at a given temperature and focusing, for example, on $\Sigma ''$ at the lowest Matsubara frequency, whose value is close to the scattering rate. After analytic continuation, these minor corrections lead to sizeable changes in real-frequency observables, as we discuss in the following sections. Since the self-energy in Matsubara frequencies is weakly affected by the presence of Landau levels, statistical errors may make it hard to see those effect when Monte-Carlo impurity solvers are used.\\

The weak effect of Landau levels on functions in Matsubara frequency can be seen from the following form of the DMFT self-consistency loop,
\begin{equation}\label{Eq.Selfconsistancy}
G^{\text{int}}(i\omega_n) = \int d\epsilon \frac{N(\epsilon)}{i\omega_n +\mu -\epsilon - \Sigma(i\omega_n)}
\end{equation}
where $N(\epsilon)$ is the non-interacting density of states. If the temperature is such that $i\omega_n$ is larger than the Landau level separation appearing in $N(\epsilon)$, their effect will be essentially washed out in $G^{\text{int}}(i\omega_n)$ by the integration over $\epsilon$. This corresponds physically to the expectation that quantum oscillations cannot be seen if temperature is much larger than Landau level separation. At lower temperature, suppose one would like to obtain information on the value at zero real frequency to detect the effect of a dimensionless magnetic flux per plaquette of order, say, 1/100 using only the first Matsubara frequency instead of the full analytic continuation. Then a value of $\beta$ of order 300 is required to obtain the same accuracy as $i\omega_n \rightarrow \omega + i\eta$ in Eq.~\eqref{Eq.Selfconsistancy} with a Lorentzian broadening $\eta = 0.01$. This value of $\beta$ is hard to reach for Monte-Carlo impurity solvers or IPT. In this case, the effect of a magnetic field might be easier to detect with a real-frequency based impurity solver such as the numerical renormalization group\cite{Bulla:2008}. 

\begin{figure}[h!]
\includegraphics[width=0.51\textwidth]{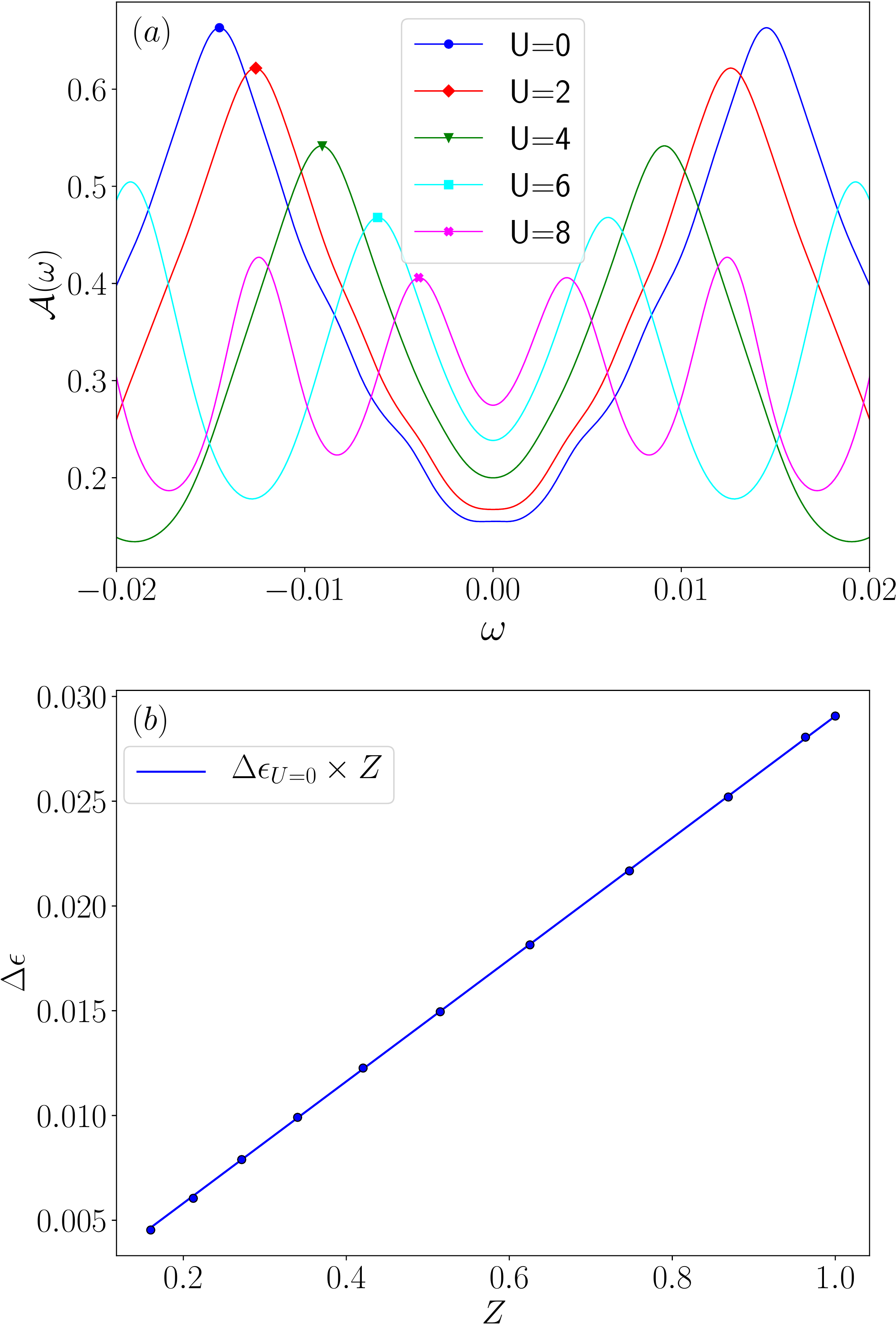} 
\centering
\caption{(Color online) (a) Density of states near Fermi energy for different values of $U$ at flux $1/q=1/100$ and $\beta = 100$. Those densities are obtained by analytic continuation using the Pad\'e approximant method\cite{Baker:1975,Vidberg_1977}. The position of the maxima is a function of $U$. (b) Blue circles: Energy separation between the two closest peaks to the Fermi energy as a function of the quasi-particle spectral weight $Z$. The continuous line is not a linear fit but the function $\Delta \epsilon_0 \times Z$ where $\Delta \epsilon_0$ is the Landau level separation at $U=0$. This shows that the cyclotron frequency is modified by the quasiparticle mass renormalization.}\label{Fig.DOS&Z}
\end{figure}  

\subsection{Effective mass from the local density of states}

Kohn's theorem\cite{Kohn_1961} states that the cyclotron resonance frequency and the de Haas-van Alphen period are independent of electron-electron interactions. However, this theorem is valid for correlation functions such as magnetization-magnetization that involve particle-hole excitations. The mass renormalization caused by interactions should be visible in the single-particle density of states, an effect that is reproduced by DMFT. It gives an alternative to the usual way of accessing the effective mass through the Lifshitz–Kosevich temperature-dependence of the amplitude of quantum oscillations. 

Fig.\ref{Fig.DOS&Z} shows, for various values of interaction strength $U$, the local density of states that can be measured by tunneling near $\omega=0$. The distance between the Landau peaks is renormalized by the change in cyclotron frequency caused by the interaction-induced mass renormalization. This can be understood as follows. The local density of states takes the form
\begin{equation}
A(\omega) = -\frac{1}{\pi} \sum_n \frac{\Sigma''(\omega)}{(\omega + \mu -\epsilon_n -\Sigma'(\omega))^2+(\Sigma''(\omega))^2},
\end{equation}
where $\epsilon_n$ labels the eigenenergies of the Harper equation. In the metallic phase one can, as usual in Fermi liquid theory, expand the real part of $\Sigma$ in power of $\omega$. This renormalizes the energy difference between Landau levels $\Delta \epsilon_{U=0}$ by the quasiparticle spectral weight $Z$: $\Delta \epsilon_{U} = Z \Delta \epsilon_{U=0}$ where, as usual,  
\begin{equation}
Z^{-1} = 1-\left. \frac{\partial  \Sigma'(\omega)}{\partial \omega}\right|_{\omega=0}.
\end{equation}
Since the self-energy is purely local, $Z$ is equivalent to the ratio between masses $m/m^*$. One can obtain $\Delta \epsilon$ by using in the expression for the cyclotron frequency the mass dressed by interactions instead of the bare band-mass of the electron. As seen in Fig.\ref{Fig.DOS&Z}, the energy of the peak nearest to the Fermi energy is modified linearly by $Z$. Note that the continuous line is not a linear fit but is directly the product $Z\Delta \epsilon_{U=0}$ with $\Delta \epsilon_{U=0}$ obtained by solving the Harper equation.

\subsection{Quantum oscillations in the lifetime}

\begin{figure}[h!]
\includegraphics[width=0.5\textwidth]{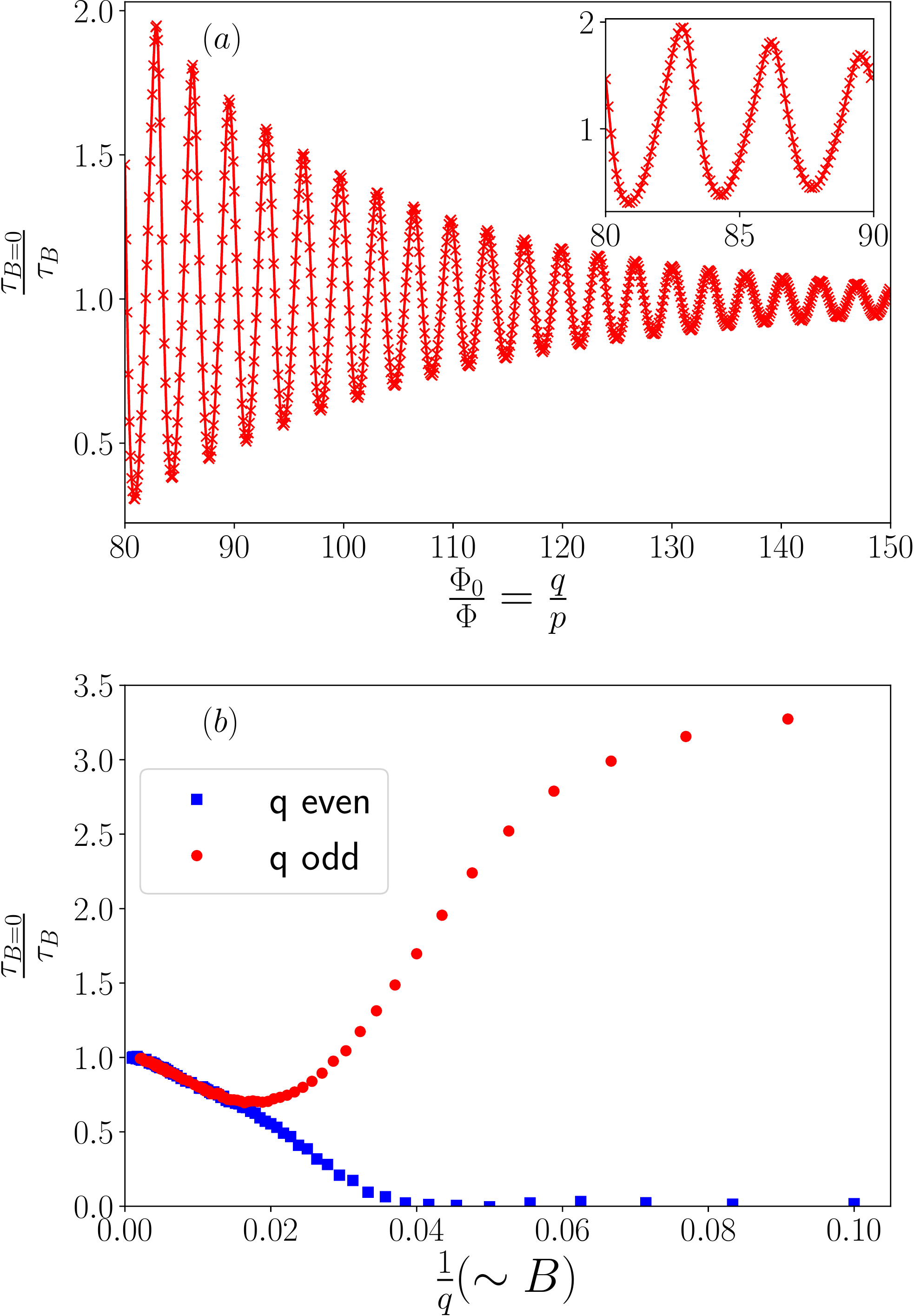}
\caption{(Color online)  (a) Scattering rate, or inverse of the  scattering time, of electrons as a function of inverse dimensionless magnetic flux $q/p$ far from half-flling ($n=0.6$) for $U=4$ and $\beta=80$. We used different values of $p$ and $q$ in order to obtain a fine grid. $q/p$ is directly proportional to $1/B$. We thus conclude that we are in the Landau regime since the period of oscillation is constant. The inset is a zoom in the high magnetic field regime that illustrates that an asymmetry between maxima and minima develops there.
		(b) Inverse of the scattering time of electrons at half-filling as a function of $1/q$ for $U=2$ and $\beta=80$. Depending on the parity of q, two different behaviors are visibles in the high-field Hofstadter's regime, \emph{i.e} a scattering time either tending to zero or to a finite value. All normalizations are with respect to the scattering time at zero flux.}\label{Fig.Demi_vies}
\end{figure}

Far from half-filling, we recover well known properties of metals subject to a uniform external magnetic field, \emph{i.e} quantum oscillations. In Fig.\ref{Fig.Demi_vies}, we show quantum oscillations of the electron's lifetime for $n = 0.6$ as a function of the inverse of the magnetic field. The period is constant in the range of magnetic field that we investigated. This is the usual behavior of observables in an electron gas and can be derived by using second order perturbation theory on the (local) self-energy and a Poisson summation formula. This constant period is a direct signature of Landau's regime. Here, we have taken different values of $p$ and $q$ in order to obtain a finer grid of magnetic-field values. We define the electron's lifetime as $\tau^{-1} = -2 Z \Sigma''(\omega = 0)$. We compute the self-energy at zero frequency using a polynomial extrapolation of the self-energy on a few of the lowest Matsubara frequencies \footnote{We checked convergence by varying the degree of the polynomial fit and checked the final result with analytic continuation using the Pad\'e method.}. At low magnetic fields (on the right of the plot), the lifetime has a nice cosine behavior which leads to Shubnikov-de Haas oscillations since the Drude conductivity is directly proportional to the scattering time. At high magnetic fields our nice sinusoidal oscillations are replaced by asymmetric periodic peaks that announce the beginning of the quantum Hall effect where the scattering rate eventually vanishes in the quantum Hall plateaus (that we do not reach here). We stress that the oscillations of $\tau$ come mainly from $\Sigma''(\omega=0)$ and,  to a lesser extent, from $Z$. The latter oscillates with the same periodicity as $\Sigma''$ but with a much smaller amplitude (around $\pm 10^{-4}$ for the same parameters as in Fig.~\ref{Fig.Demi_vies}. The imaginary part of the self-energy is much more sensitive to Fermi-surface effects than $Z$, which depends on the real part of the self-energy, hence on virtual processes on many energy scales (as follows from Kramers-Kronig). 

Striking differences between the high- and low-field cases occur when the Hofstadter butterfly regime becomes visible. One of the most remarkable features of this regime is the difference between even and odd values of $q$ when $\Phi/\Phi_0=1/q$. For $q$ odd, the system is in a metallic phase in the sense that it has a finite density of states at the Fermi energy. For $q$ even, the system is in a semimetallic phase with $q$ non-equivalent Dirac cones in the magnetic Brillouin zone~\cite{Kohmoto_1989}. This property is directly visible in the scattering rate of electrons. This is illustrated in Fig.~\ref{Fig.Demi_vies} where we plot $\tau_{B=0}/\tau_B$ as a function of magnetic flux at $U=2$ and $\beta=80$. For low magnetic fields compared to the thermal energy, the system makes no differences between odd and even values of $q$. When $1/q$ is of the order of $1/\beta$, details of the density of states at the Fermi energy become noticeable, leading to two different tendencies, \emph{i.e.} a scattering rate tending either to zero or to a finite value depending in the parity of $q$. At low field one would expect a plateau when thermal effects wash out simple Landau levels arising from a density of states that is constant at $B=0$. Here, this argument does not work out possibly because of the presence of a Van-Hove singularity at half-filling on the square lattice~\cite{Naumis_2016}.\\

\section{Conclusion}\label{Sec:Conclusion} 

We have shown how to include the orbital effect of a uniform magnetic field within dynamical mean-field theory using the cavity method. The self-consistency relation is not modified. The magnetic field comes in through the non-interacting band structure and a single translationally-invariant impurity problem needs to be solved. This result has already been used without rigorous proof to study various problems such as the Hofstadter butterfly or the Falicov-Kimball model. The advantage of DMFT compared to RDMFT lies in the reduction in needed computational power since one has to solve only one quantum impurity problem. 

As an example of application, we used the DMFT method to recover effects of the magnetic field on interacting lattice electrons. The scattering time and the density of states are two measurable properties that are affected by the combined effect of magnetic field and interactions. The scattering time shows quantum oscillations and has a non-trivial dependence on the magnetic field that can vary according to the filling and the amplitude of $B$, while the cyclotron frequency observable in the density of states is modified because of the interaction-induced effective mass. Quantum oscillations in the effective mass are negligibly small. In the Hofstadter regime, the density of states and the self-energy are both strongly affected by the combined effects of  magnetic field and interactions. It would be interesting to verify the results depicted in Fig.~\ref{Fig.Demi_vies} in cold-atom experiments.  

\begin{acknowledgments}
We thank M. Berciu, M. Charlebois and R. Nourafkan for useful discussions. This work has been supported by the Natural Sciences and Engineering Research Council of Canada (NSERC) under grant RGPIN-2014-04584, and by the Research Chair in the Theory of Quantum Materials. Simulations were performed on computers provided by the Canadian Foundation for Innovation, the Minist\`ere de l'\'Education des Loisirs et du Sport (Qu\'ebec), Calcul Qu\'ebec, and Compute Canada.
\end{acknowledgments}

\appendix 
\section{Asymptotic behavior of the local Green function and hybridization}\label{Appendix:asymptotics}

Start from the expression for $\mathcal{G}_0$ given by the self-consistency equation \eqref{{Eq.Weiss_final_dmft}}. Using translational invariance, rewrite $G_{ll}(i\omega_n)$ on the right-hand side in the following site representation~\cite{Arsenault:these}
\begin{align}
G_{ll}(i\omega_n)&=\frac{1}{N}\sum_{\l}^{N}G_{ll}(i\omega_n)={\rm Tr}[G(i\omega_n)]\\
&={\rm Tr}\left[[(i\omega_n+\mu-\Sigma(i\omega_n))\mathbf{I}-\mathbf{H^0}]^{-1}\right]
\end{align}
where $\mathbf{I}$ is the identity matrix and $\mathbf{H^0}$ the Hamiltonian matrix whose expression can be deduced from Eq.~(\ref{Eq:Hamiltonian}) with the interaction and chemical potential terms removed. Expanding the above equation to leading order in $(i\omega_n)^{-2}$, we find~\cite{Koch:2008}
\begin{equation}
G_{ll}(i\omega_n)\simeq \frac{1}{i\omega_n}{\rm Tr}\left[\mathbf{I}-\frac{\mathbf{X}}{i\omega_n}+\frac{\mathbf{X}\mathbf{X}}{(i\omega_n)^2}+\cdots\right]\, .
\label{Eq:Glocal}
\end{equation}
with 
\begin{equation}
\mathbf{X}=(\mu-\Sigma(i\omega_n))\mathbf{I}-\mathbf{H^0}\, .
\end{equation}
Inverting, expanding again and substituting in the self-consistency equation \eqref{Eq.Green_l} we find, whatever the asymptotic behavior of the self-energy, the following expression for the aymptotic behavior of the hybridization function
\begin{equation}
\Delta(i\omega_n)\simeq \frac{{\rm Tr}[\mathbf{H^0}\mathbf{H^0}]-{\rm Tr}[\mathbf{H^0}]^2}{i\omega_n}
\end{equation}
which should be used in the definition of the asymptotic form of $\mathcal{G}_0$ in Eq.~(\ref{Eq:Weiss}).
%\begin{equation} 
%\mathcal{G}^{\text{inf}}_0(i\omega_n) = \frac{1}{i\omega_n +\mu - {\rm %Tr}[\mathbf{H^0}] - \frac{c}{i\omega_m}}
%\end{equation}
%with the definition
%\begin{equation}
%c={\rm Tr}[\mathbf{H^0}\mathbf{H^0}]-{\rm Tr}[\mathbf{H^0}]^2\, .
%\end{equation}
Given that the Peierls phase vanishes in the diagonal elements of $\mathbf{H^0}$ and that it changes sign when $\mathbf{H^0}$ is transposed, both ${\rm Tr}[\mathbf{H^0}]$ and ${\rm Tr}[\mathbf{H^0}\mathbf{H^0}]=\sum_{i.j}H^0_{ij}H^0_{ji}/N$ are independent of magnetic field~\cite{Arsenault:these} and can be easily calculated in the diagonal basis for $\mathbf{H^0}$.  Note that ${\rm Tr}[\mathbf{H^0}]=0$ in our case.

The complete asymptotic behavior of the local Green function $G_{ll}(i\omega_n)$ can be found using the usual procedure of expanding the spectral representation up to order $(i\omega_n)^{-3}$, which leads to coefficients of the successive terms that are expressed as moments of the spectral function. These moments are in turn obtained from equal-time commutators. An equivalent procedure starting from  Eq.~\eqref{Eq:Glocal} leads to ${\rm Tr}[\mathbf{H^0}]$ and ${\rm Tr}[\mathbf{H^0}\mathbf{H^0}]$ terms whose Peierls phase disappears and also to terms that depend on the expansion of $\Sigma(i\omega_n)$ to order $(i\omega_n)^{-1}$, which is found using the equal-time commutator procedure mentioned above. It is found that the expansion of $\Sigma(i\omega_n)$ up to order $(i\omega_n)^{-1}$ does not depend on $\mathbf{B}$~\cite{Arsenault:these}. This is not surprising since the expansion of $\Sigma(i\omega_n)$ up to order $(i\omega_n)^{-1}$  in the $\mathbf{B}=\mathbf{0}$ case depends only on $U$ and on occupation number. So, finally, the expansion of $G_{ll}(i\omega_n)$ up to order $(i\omega_n)^{-3}$ does not depend on $\mathbf{B}$.

%\bibliography{Bibliography}

\begin{thebibliography}{38}%
	\makeatletter
	\providecommand \@ifxundefined [1]{%
		\@ifx{#1\undefined}
	}%
	\providecommand \@ifnum [1]{%
		\ifnum #1\expandafter \@firstoftwo
		\else \expandafter \@secondoftwo
		\fi
	}%
	\providecommand \@ifx [1]{%
		\ifx #1\expandafter \@firstoftwo
		\else \expandafter \@secondoftwo
		\fi
	}%
	\providecommand \natexlab [1]{#1}%
	\providecommand \enquote  [1]{``#1''}%
	\providecommand \bibnamefont  [1]{#1}%
	\providecommand \bibfnamefont [1]{#1}%
	\providecommand \citenamefont [1]{#1}%
	\providecommand \href@noop [0]{\@secondoftwo}%
	\providecommand \href [0]{\begingroup \@sanitize@url \@href}%
	\providecommand \@href[1]{\@@startlink{#1}\@@href}%
	\providecommand \@@href[1]{\endgroup#1\@@endlink}%
	\providecommand \@sanitize@url [0]{\catcode `\\12\catcode `\$12\catcode
		`\&12\catcode `\#12\catcode `\^12\catcode `\_12\catcode `\%12\relax}%
	\providecommand \@@startlink[1]{}%
	\providecommand \@@endlink[0]{}%
	\providecommand \url  [0]{\begingroup\@sanitize@url \@url }%
	\providecommand \@url [1]{\endgroup\@href {#1}{\urlprefix }}%
	\providecommand \urlprefix  [0]{URL }%
	\providecommand \Eprint [0]{\href }%
	\providecommand \doibase [0]{http://dx.doi.org/}%
	\providecommand \selectlanguage [0]{\@gobble}%
	\providecommand \bibinfo  [0]{\@secondoftwo}%
	\providecommand \bibfield  [0]{\@secondoftwo}%
	\providecommand \translation [1]{[#1]}%
	\providecommand \BibitemOpen [0]{}%
	\providecommand \bibitemStop [0]{}%
	\providecommand \bibitemNoStop [0]{.\EOS\space}%
	\providecommand \EOS [0]{\spacefactor3000\relax}%
	\providecommand \BibitemShut  [1]{\csname bibitem#1\endcsname}%
	\let\auto@bib@innerbib\@empty
	%</preamble>
	\bibitem [{\citenamefont {Georges}\ and\ \citenamefont
		{Kotliar}(1992)}]{Georges_1992}%
	\BibitemOpen
	\bibfield  {author} {\bibinfo {author} {\bibfnamefont {Antoine}\ \bibnamefont
			{Georges}}\ and\ \bibinfo {author} {\bibfnamefont {Gabriel}\ \bibnamefont
			{Kotliar}},\ }\bibfield  {title} {\enquote {\bibinfo {title} {Hubbard model
				in infinite dimensions},}\ }\href {\doibase 10.1103/PhysRevB.45.6479}
	{\bibfield  {journal} {\bibinfo  {journal} {Phys. Rev. B}\ }\textbf {\bibinfo
			{volume} {45}},\ \bibinfo {pages} {6479--6483} (\bibinfo {year}
		{1992})}\BibitemShut {NoStop}%
	\bibitem [{\citenamefont {Georges}\ \emph {et~al.}(1996)\citenamefont
		{Georges}, \citenamefont {Kotliar}, \citenamefont {Krauth},\ and\
		\citenamefont {Rozenberg}}]{Georges_1996}%
	\BibitemOpen
	\bibfield  {author} {\bibinfo {author} {\bibfnamefont {Antoine}\ \bibnamefont
			{Georges}}, \bibinfo {author} {\bibfnamefont {Gabriel}\ \bibnamefont
			{Kotliar}}, \bibinfo {author} {\bibfnamefont {Werner}\ \bibnamefont
			{Krauth}}, \ and\ \bibinfo {author} {\bibfnamefont {Marcelo~J.}\ \bibnamefont
			{Rozenberg}},\ }\bibfield  {title} {\enquote {\bibinfo {title} {Dynamical
				mean-field theory of strongly correlated fermion systems and the limit of
				infinite dimensions},}\ }\href {\doibase 10.1103/RevModPhys.68.13} {\bibfield
		{journal} {\bibinfo  {journal} {Rev. Mod. Phys.}\ }\textbf {\bibinfo
			{volume} {68}},\ \bibinfo {pages} {13--125} (\bibinfo {year}
		{1996})}\BibitemShut {NoStop}%
	\bibitem [{\citenamefont {Hofstadter}(1976)}]{Hofstadter_1976}%
	\BibitemOpen
	\bibfield  {author} {\bibinfo {author} {\bibfnamefont {Douglas~R.}\
			\bibnamefont {Hofstadter}},\ }\bibfield  {title} {\enquote {\bibinfo {title}
			{Energy levels and wave functions of bloch electrons in rational and
				irrational magnetic fields},}\ }\href {\doibase 10.1103/PhysRevB.14.2239}
	{\bibfield  {journal} {\bibinfo  {journal} {Phys. Rev. B}\ }\textbf {\bibinfo
			{volume} {14}},\ \bibinfo {pages} {2239--2249} (\bibinfo {year}
		{1976})}\BibitemShut {NoStop}%
	\bibitem [{\citenamefont {MacDonald}(1983)}]{MacDonald_1983}%
	\BibitemOpen
	\bibfield  {author} {\bibinfo {author} {\bibfnamefont {A.~H.}\ \bibnamefont
			{MacDonald}},\ }\bibfield  {title} {\enquote {\bibinfo {title} {Landau-level
				subband structure of electrons on a square lattice},}\ }\href {\doibase
		10.1103/PhysRevB.28.6713} {\bibfield  {journal} {\bibinfo  {journal} {Phys.
				Rev. B}\ }\textbf {\bibinfo {volume} {28}},\ \bibinfo {pages} {6713--6717}
		(\bibinfo {year} {1983})}\BibitemShut {NoStop}%
	\bibitem [{\citenamefont {Goldman}\ \emph {et~al.}(2010)\citenamefont
		{Goldman}, \citenamefont {Satija}, \citenamefont {Nikolic}, \citenamefont
		{Bermudez}, \citenamefont {Martin-Delgado}, \citenamefont {Lewenstein},\ and\
		\citenamefont {Spielman}}]{Goldman_2010}%
	\BibitemOpen
	\bibfield  {author} {\bibinfo {author} {\bibfnamefont {N.}~\bibnamefont
			{Goldman}}, \bibinfo {author} {\bibfnamefont {I.}~\bibnamefont {Satija}},
		\bibinfo {author} {\bibfnamefont {P.}~\bibnamefont {Nikolic}}, \bibinfo
		{author} {\bibfnamefont {A.}~\bibnamefont {Bermudez}}, \bibinfo {author}
		{\bibfnamefont {M.~A.}\ \bibnamefont {Martin-Delgado}}, \bibinfo {author}
		{\bibfnamefont {M.}~\bibnamefont {Lewenstein}}, \ and\ \bibinfo {author}
		{\bibfnamefont {I.~B.}\ \bibnamefont {Spielman}},\ }\bibfield  {title}
	{\enquote {\bibinfo {title} {Realistic time-reversal invariant topological
				insulators with neutral atoms},}\ }\href {\doibase
		10.1103/PhysRevLett.105.255302} {\bibfield  {journal} {\bibinfo  {journal}
			{Phys. Rev. Lett.}\ }\textbf {\bibinfo {volume} {105}},\ \bibinfo {pages}
		{255302} (\bibinfo {year} {2010})}\BibitemShut {NoStop}%
	\bibitem [{\citenamefont {Gerbier}\ and\ \citenamefont
		{Dalibard}(2010)}]{Gerbier_2010}%
	\BibitemOpen
	\bibfield  {author} {\bibinfo {author} {\bibfnamefont {Fabrice}\ \bibnamefont
			{Gerbier}}\ and\ \bibinfo {author} {\bibfnamefont {Jean}\ \bibnamefont
			{Dalibard}},\ }\bibfield  {title} {\enquote {\bibinfo {title} {Gauge fields
				for ultracold atoms in optical superlattices},}\ }\href
	{http://stacks.iop.org/1367-2630/12/i=3/a=033007} {\bibfield  {journal}
		{\bibinfo  {journal} {New Journal of Physics}\ }\textbf {\bibinfo {volume}
			{12}},\ \bibinfo {pages} {033007} (\bibinfo {year} {2010})}\BibitemShut
	{NoStop}%
	\bibitem [{\citenamefont {Dean}\ \emph {et~al.}()\citenamefont {Dean},
		\citenamefont {Wang}, \citenamefont {Maher}, \citenamefont {Forsythe},
		\citenamefont {Ghahari}, \citenamefont {Gao}, \citenamefont {Katoch},
		\citenamefont {Ishigami}, \citenamefont {Moon}, \citenamefont {Koshino},
		\citenamefont {Taniguchi}, \citenamefont {Watanabe}, \citenamefont {Shepard},
		\citenamefont {Hone},\ and\ \citenamefont {Kim}}]{Dean_2013}%
	\BibitemOpen
	\bibfield  {author} {\bibinfo {author} {\bibfnamefont {C.~R.}\ \bibnamefont
			{Dean}}, \bibinfo {author} {\bibfnamefont {L.}~\bibnamefont {Wang}}, \bibinfo
		{author} {\bibfnamefont {P.}~\bibnamefont {Maher}}, \bibinfo {author}
		{\bibfnamefont {C.}~\bibnamefont {Forsythe}}, \bibinfo {author}
		{\bibfnamefont {F.}~\bibnamefont {Ghahari}}, \bibinfo {author} {\bibfnamefont
			{Y.}~\bibnamefont {Gao}}, \bibinfo {author} {\bibfnamefont {J.}~\bibnamefont
			{Katoch}}, \bibinfo {author} {\bibfnamefont {M.}~\bibnamefont {Ishigami}},
		\bibinfo {author} {\bibfnamefont {P.}~\bibnamefont {Moon}}, \bibinfo {author}
		{\bibfnamefont {M.}~\bibnamefont {Koshino}}, \bibinfo {author} {\bibfnamefont
			{T.}~\bibnamefont {Taniguchi}}, \bibinfo {author} {\bibfnamefont
			{K.}~\bibnamefont {Watanabe}}, \bibinfo {author} {\bibfnamefont {K.~L.}\
			\bibnamefont {Shepard}}, \bibinfo {author} {\bibfnamefont {J.}~\bibnamefont
			{Hone}}, \ and\ \bibinfo {author} {\bibfnamefont {P.}~\bibnamefont {Kim}},\
	}\bibfield  {title} {\enquote {\bibinfo {title} {Hofstadter/'s butterfly and
				the fractal quantum hall effect in moire superlattices},}\ }\href
	{http://dx.doi.org/10.1038/nature12186} {\ \textbf {\bibinfo {volume}
			{497}},\ \bibinfo {pages} {598--602}}\BibitemShut {NoStop}%
	\bibitem [{\citenamefont {Gudmundsson}\ and\ \citenamefont
		{Gerhardts}(1995)}]{Gudmundsson_1995}%
	\BibitemOpen
	\bibfield  {author} {\bibinfo {author} {\bibfnamefont {Vidar}\ \bibnamefont
			{Gudmundsson}}\ and\ \bibinfo {author} {\bibfnamefont {Rolf~R.}\ \bibnamefont
			{Gerhardts}},\ }\bibfield  {title} {\enquote {\bibinfo {title} {Effects of
				screening on the hofstadter butterfly},}\ }\href {\doibase
		10.1103/PhysRevB.52.16744} {\bibfield  {journal} {\bibinfo  {journal} {Phys.
				Rev. B}\ }\textbf {\bibinfo {volume} {52}},\ \bibinfo {pages} {16744--16752}
		(\bibinfo {year} {1995})}\BibitemShut {NoStop}%
	\bibitem [{\citenamefont {Mishra}\ \emph {et~al.}(2016)\citenamefont {Mishra},
		\citenamefont {Hassan},\ and\ \citenamefont {Shankar}}]{Mishra_2016}%
	\BibitemOpen
	\bibfield  {author} {\bibinfo {author} {\bibfnamefont {Archana}\ \bibnamefont
			{Mishra}}, \bibinfo {author} {\bibfnamefont {S.~R.}\ \bibnamefont {Hassan}},
		\ and\ \bibinfo {author} {\bibfnamefont {R.}~\bibnamefont {Shankar}},\
	}\bibfield  {title} {\enquote {\bibinfo {title} {Effects of interaction in
				the hofstadter regime of the honeycomb lattice},}\ }\href {\doibase
		10.1103/PhysRevB.93.125134} {\bibfield  {journal} {\bibinfo  {journal} {Phys.
				Rev. B}\ }\textbf {\bibinfo {volume} {93}},\ \bibinfo {pages} {125134}
		(\bibinfo {year} {2016})}\BibitemShut {NoStop}%
	\bibitem [{\citenamefont {Doh}\ and\ \citenamefont {Salk}(1998)}]{Doh_1998}%
	\BibitemOpen
	\bibfield  {author} {\bibinfo {author} {\bibfnamefont {Hyeonjin}\
			\bibnamefont {Doh}}\ and\ \bibinfo {author} {\bibfnamefont {Sung-Ho~Suck}\
			\bibnamefont {Salk}},\ }\bibfield  {title} {\enquote {\bibinfo {title}
			{Effects of electron correlations on the hofstadter spectrum},}\ }\href
	{\doibase 10.1103/PhysRevB.57.1312} {\bibfield  {journal} {\bibinfo
			{journal} {Phys. Rev. B}\ }\textbf {\bibinfo {volume} {57}},\ \bibinfo
		{pages} {1312--1315} (\bibinfo {year} {1998})}\BibitemShut {NoStop}%
	\bibitem [{\citenamefont {Tran}(2010)}]{Tran_2010}%
	\BibitemOpen
	\bibfield  {author} {\bibinfo {author} {\bibfnamefont {Minh-Tien}\
			\bibnamefont {Tran}},\ }\bibfield  {title} {\enquote {\bibinfo {title}
			{Electronic structure of the falicov-kimball model with a magnetic field:
				Dynamical mean-field study},}\ }\href {\doibase 10.1103/PhysRevB.81.115119}
	{\bibfield  {journal} {\bibinfo  {journal} {Phys. Rev. B}\ }\textbf {\bibinfo
			{volume} {81}},\ \bibinfo {pages} {115119} (\bibinfo {year}
		{2010})}\BibitemShut {NoStop}%
	\bibitem [{\citenamefont {Cocks}\ \emph {et~al.}(2012)\citenamefont {Cocks},
		\citenamefont {Orth}, \citenamefont {Rachel}, \citenamefont {Buchhold},
		\citenamefont {Le~Hur},\ and\ \citenamefont {Hofstetter}}]{Cocks_2012}%
	\BibitemOpen
	\bibfield  {author} {\bibinfo {author} {\bibfnamefont {Daniel}\ \bibnamefont
			{Cocks}}, \bibinfo {author} {\bibfnamefont {Peter~P.}\ \bibnamefont {Orth}},
		\bibinfo {author} {\bibfnamefont {Stephan}\ \bibnamefont {Rachel}}, \bibinfo
		{author} {\bibfnamefont {Michael}\ \bibnamefont {Buchhold}}, \bibinfo
		{author} {\bibfnamefont {Karyn}\ \bibnamefont {Le~Hur}}, \ and\ \bibinfo
		{author} {\bibfnamefont {Walter}\ \bibnamefont {Hofstetter}},\ }\bibfield
	{title} {\enquote {\bibinfo {title} {Time-reversal-invariant
				hofstadter-hubbard model with ultracold fermions},}\ }\href {\doibase
		10.1103/PhysRevLett.109.205303} {\bibfield  {journal} {\bibinfo  {journal}
			{Phys. Rev. Lett.}\ }\textbf {\bibinfo {volume} {109}},\ \bibinfo {pages}
		{205303} (\bibinfo {year} {2012})}\BibitemShut {NoStop}%
	\bibitem [{\citenamefont {Kumar}\ \emph {et~al.}(2016)\citenamefont {Kumar},
		\citenamefont {Mertz},\ and\ \citenamefont {Hofstetter}}]{Kumar_2016}%
	\BibitemOpen
	\bibfield  {author} {\bibinfo {author} {\bibfnamefont {Pramod}\ \bibnamefont
			{Kumar}}, \bibinfo {author} {\bibfnamefont {Thomas}\ \bibnamefont {Mertz}}, \
		and\ \bibinfo {author} {\bibfnamefont {Walter}\ \bibnamefont {Hofstetter}},\
	}\bibfield  {title} {\enquote {\bibinfo {title} {Interaction-induced
				topological and magnetic phases in the hofstadter-hubbard model},}\ }\href
	{\doibase 10.1103/PhysRevB.94.115161} {\bibfield  {journal} {\bibinfo
			{journal} {Phys. Rev. B}\ }\textbf {\bibinfo {volume} {94}},\ \bibinfo
		{pages} {115161} (\bibinfo {year} {2016})}\BibitemShut {NoStop}%
	\bibitem [{\citenamefont {Orth}\ \emph {et~al.}(2013)\citenamefont {Orth},
		\citenamefont {Cocks}, \citenamefont {Rachel}, \citenamefont {Buchhold},
		\citenamefont {Hur},\ and\ \citenamefont {Hofstetter}}]{Orth_2013}%
	\BibitemOpen
	\bibfield  {author} {\bibinfo {author} {\bibfnamefont {Peter~P}\ \bibnamefont
			{Orth}}, \bibinfo {author} {\bibfnamefont {Daniel}\ \bibnamefont {Cocks}},
		\bibinfo {author} {\bibfnamefont {Stephan}\ \bibnamefont {Rachel}}, \bibinfo
		{author} {\bibfnamefont {Michael}\ \bibnamefont {Buchhold}}, \bibinfo
		{author} {\bibfnamefont {Karyn~Le}\ \bibnamefont {Hur}}, \ and\ \bibinfo
		{author} {\bibfnamefont {Walter}\ \bibnamefont {Hofstetter}},\ }\bibfield
	{title} {\enquote {\bibinfo {title} {Correlated topological phases and exotic
				magnetism with ultracold fermions},}\ }\href
	{http://stacks.iop.org/0953-4075/46/i=13/a=134004} {\bibfield  {journal}
		{\bibinfo  {journal} {Journal of Physics B: Atomic, Molecular and Optical
				Physics}\ }\textbf {\bibinfo {volume} {46}},\ \bibinfo {pages} {134004}
		(\bibinfo {year} {2013})}\BibitemShut {NoStop}%
	\bibitem [{\citenamefont {Potthoff}\ and\ \citenamefont
		{Nolting}(1999)}]{Potthoff_1999}%
	\BibitemOpen
	\bibfield  {author} {\bibinfo {author} {\bibfnamefont {M.}~\bibnamefont
			{Potthoff}}\ and\ \bibinfo {author} {\bibfnamefont {W.}~\bibnamefont
			{Nolting}},\ }\bibfield  {title} {\enquote {\bibinfo {title} {Surface
				metal-insulator transition in the hubbard model},}\ }\href {\doibase
		10.1103/PhysRevB.59.2549} {\bibfield  {journal} {\bibinfo  {journal} {Phys.
				Rev. B}\ }\textbf {\bibinfo {volume} {59}},\ \bibinfo {pages} {2549--2555}
		(\bibinfo {year} {1999})}\BibitemShut {NoStop}%
	\bibitem [{\citenamefont {Snoek}\ \emph {et~al.}(2008)\citenamefont {Snoek},
		\citenamefont {Titvinidze}, \citenamefont {Tőke}, \citenamefont {Byczuk},\
		and\ \citenamefont {Hofstetter}}]{Snoek_2008}%
	\BibitemOpen
	\bibfield  {author} {\bibinfo {author} {\bibfnamefont {M}~\bibnamefont
			{Snoek}}, \bibinfo {author} {\bibfnamefont {I}~\bibnamefont {Titvinidze}},
		\bibinfo {author} {\bibfnamefont {C}~\bibnamefont {Tőke}}, \bibinfo {author}
		{\bibfnamefont {K}~\bibnamefont {Byczuk}}, \ and\ \bibinfo {author}
		{\bibfnamefont {W}~\bibnamefont {Hofstetter}},\ }\bibfield  {title} {\enquote
		{\bibinfo {title} {Antiferromagnetic order of strongly interacting fermions
				in a trap: real-space dynamical mean-field analysis},}\ }\href
	{http://stacks.iop.org/1367-2630/10/i=9/a=093008} {\bibfield  {journal}
		{\bibinfo  {journal} {New Journal of Physics}\ }\textbf {\bibinfo {volume}
			{10}},\ \bibinfo {pages} {093008} (\bibinfo {year} {2008})}\BibitemShut
	{NoStop}%
	\bibitem [{\citenamefont {Arsenault}(2013)}]{Arsenault:these}%
	\BibitemOpen
	\bibfield  {author} {\bibinfo {author} {\bibfnamefont {Louis-François}\
			\bibnamefont {Arsenault}},\ }\emph {\bibinfo {title} {Nouvelles approches en
			th\'eorie du champ moyen dynamique : le cas du pouvoir thermo\'electrique et
			celui de l'effet orbital d'un champ magn\'etique}},\ \href@noop {} {Ph.D.
		thesis},\ \bibinfo  {school} {Universit\'e de Sherbrooke} (\bibinfo {year}
	{2013})\BibitemShut {NoStop}%
	\bibitem [{\citenamefont {Laloux}\ \emph {et~al.}(1994)\citenamefont {Laloux},
		\citenamefont {Georges},\ and\ \citenamefont {Krauth}}]{LalouxGeorges:1994}%
	\BibitemOpen
	\bibfield  {author} {\bibinfo {author} {\bibfnamefont {Laurent}\ \bibnamefont
			{Laloux}}, \bibinfo {author} {\bibfnamefont {Antoine}\ \bibnamefont
			{Georges}}, \ and\ \bibinfo {author} {\bibfnamefont {Werner}\ \bibnamefont
			{Krauth}},\ }\bibfield  {title} {\enquote {\bibinfo {title} {Effect of a
				magnetic field on mott-hubbard systems},}\ }\href {\doibase
		10.1103/PhysRevB.50.3092} {\bibfield  {journal} {\bibinfo  {journal} {Phys.
				Rev. B}\ }\textbf {\bibinfo {volume} {50}},\ \bibinfo {pages} {3092--3102}
		(\bibinfo {year} {1994})}\BibitemShut {NoStop}%
	\bibitem [{\citenamefont {Kita}\ and\ \citenamefont
		{Arai}(2005)}]{Takafumi:2005}%
	\BibitemOpen
	\bibfield  {author} {\bibinfo {author} {\bibfnamefont {Takafumi}\
			\bibnamefont {Kita}}\ and\ \bibinfo {author} {\bibfnamefont {Masao}\
			\bibnamefont {Arai}},\ }\bibfield  {title} {\enquote {\bibinfo {title}
			{Theory of interacting bloch electrons in a magnetic field},}\ }\href
	{\doibase 10.1143/JPSJ.74.2813} {\bibfield  {journal} {\bibinfo  {journal}
			{Journal of the Physical Society of Japan}\ }\textbf {\bibinfo {volume}
			{74}},\ \bibinfo {pages} {2813--2830} (\bibinfo {year} {2005})},\ \Eprint
	{http://arxiv.org/abs/http://dx.doi.org/10.1143/JPSJ.74.2813}
	{http://dx.doi.org/10.1143/JPSJ.74.2813} \BibitemShut {NoStop}%
	\bibitem [{Note1()}]{Note1}%
	\BibitemOpen
	\bibinfo {note} {We checked convergence by varying the degree of the
		polynomial fit and checked the final result with analytic continuation using
		the Pad\'e method.}\BibitemShut {Stop}%
	\bibitem [{\citenamefont {Hubbard}(1964)}]{Hubbard:1964}%
	\BibitemOpen
	\bibfield  {author} {\bibinfo {author} {\bibfnamefont {J.}~\bibnamefont
			{Hubbard}},\ }\href@noop {} {\bibfield  {journal} {\bibinfo  {journal} {Proc.
				Roy. Soc. (London)}\ }\textbf {\bibinfo {volume} {A 281}},\ \bibinfo {pages}
		{, 401.} (\bibinfo {year} {1964})}\BibitemShut {NoStop}%
	\bibitem [{\citenamefont {Chen}\ and\ \citenamefont {Lee}(2011)}]{Chen_2011}%
	\BibitemOpen
	\bibfield  {author} {\bibinfo {author} {\bibfnamefont {Kuang-Ting}\
			\bibnamefont {Chen}}\ and\ \bibinfo {author} {\bibfnamefont {Patrick~A.}\
			\bibnamefont {Lee}},\ }\bibfield  {title} {\enquote {\bibinfo {title}
			{Unified formalism for calculating polarization, magnetization, and more in a
				periodic insulator},}\ }\href {\doibase 10.1103/PhysRevB.84.205137}
	{\bibfield  {journal} {\bibinfo  {journal} {Phys. Rev. B}\ }\textbf {\bibinfo
			{volume} {84}},\ \bibinfo {pages} {205137} (\bibinfo {year}
		{2011})}\BibitemShut {NoStop}%
	\bibitem [{\citenamefont {Khodas}\ and\ \citenamefont
		{Finkel'stein}(2003)}]{Khodas_2003}%
	\BibitemOpen
	\bibfield  {author} {\bibinfo {author} {\bibfnamefont {M.}~\bibnamefont
			{Khodas}}\ and\ \bibinfo {author} {\bibfnamefont {A.~M.}\ \bibnamefont
			{Finkel'stein}},\ }\bibfield  {title} {\enquote {\bibinfo {title} {Hall
				coefficient in an interacting electron gas},}\ }\href {\doibase
		10.1103/PhysRevB.68.155114} {\bibfield  {journal} {\bibinfo  {journal} {Phys.
				Rev. B}\ }\textbf {\bibinfo {volume} {68}},\ \bibinfo {pages} {155114}
		(\bibinfo {year} {2003})}\BibitemShut {NoStop}%
	\bibitem [{\citenamefont {Berciu}\ and\ \citenamefont
		{Cook}(2010)}]{Berciu_2010}%
	\BibitemOpen
	\bibfield  {author} {\bibinfo {author} {\bibfnamefont {M.}~\bibnamefont
			{Berciu}}\ and\ \bibinfo {author} {\bibfnamefont {A.~M.}\ \bibnamefont
			{Cook}},\ }\bibfield  {title} {\enquote {\bibinfo {title} {Efficient
				computation of lattice green's functions for models with nearest-neighbour
				hopping},}\ }\href {http://stacks.iop.org/0295-5075/92/i=4/a=40003}
	{\bibfield  {journal} {\bibinfo  {journal} {EPL (Europhysics Letters)}\
		}\textbf {\bibinfo {volume} {92}},\ \bibinfo {pages} {40003} (\bibinfo {year}
		{2010})}\BibitemShut {NoStop}%
	\bibitem [{\citenamefont {Ueta}(1997)}]{Ueta_1997}%
	\BibitemOpen
	\bibfield  {author} {\bibinfo {author} {\bibfnamefont {Tsuyoshi}\
			\bibnamefont {Ueta}},\ }\bibfield  {title} {\enquote {\bibinfo {title}
			{Lattice green function in uniform magnetic fields},}\ }\href
	{http://stacks.iop.org/0305-4470/30/i=15/a=020} {\bibfield  {journal}
		{\bibinfo  {journal} {Journal of Physics A: Mathematical and General}\
		}\textbf {\bibinfo {volume} {30}},\ \bibinfo {pages} {5339} (\bibinfo {year}
		{1997})}\BibitemShut {NoStop}%
	\bibitem [{\citenamefont {Harper}(1955)}]{Harper_1955}%
	\BibitemOpen
	\bibfield  {author} {\bibinfo {author} {\bibfnamefont {P~G}\ \bibnamefont
			{Harper}},\ }\bibfield  {title} {\enquote {\bibinfo {title} {Single band
				motion of conduction electrons in a uniform magnetic field},}\ }\href
	{http://stacks.iop.org/0370-1298/68/i=10/a=304} {\bibfield  {journal}
		{\bibinfo  {journal} {Proceedings of the Physical Society. Section A}\
		}\textbf {\bibinfo {volume} {68}},\ \bibinfo {pages} {874} (\bibinfo {year}
		{1955})}\BibitemShut {NoStop}%
	\bibitem [{\citenamefont {Kajueter}\ and\ \citenamefont
		{Kotliar}(1996)}]{Kajueter_1996}%
	\BibitemOpen
	\bibfield  {author} {\bibinfo {author} {\bibfnamefont {Henrik}\ \bibnamefont
			{Kajueter}}\ and\ \bibinfo {author} {\bibfnamefont {Gabriel}\ \bibnamefont
			{Kotliar}},\ }\bibfield  {title} {\enquote {\bibinfo {title} {New iterative
				perturbation scheme for lattice models with arbitrary filling},}\ }\href
	{\doibase 10.1103/PhysRevLett.77.131} {\bibfield  {journal} {\bibinfo
			{journal} {Phys. Rev. Lett.}\ }\textbf {\bibinfo {volume} {77}},\ \bibinfo
		{pages} {131--134} (\bibinfo {year} {1996})}\BibitemShut {NoStop}%
	\bibitem [{\citenamefont {Zhang}\ \emph {et~al.}(1993)\citenamefont {Zhang},
		\citenamefont {Rozenberg},\ and\ \citenamefont {Kotliar}}]{Zhang_1993}%
	\BibitemOpen
	\bibfield  {author} {\bibinfo {author} {\bibfnamefont {X.~Y.}\ \bibnamefont
			{Zhang}}, \bibinfo {author} {\bibfnamefont {M.~J.}\ \bibnamefont
			{Rozenberg}}, \ and\ \bibinfo {author} {\bibfnamefont {G.}~\bibnamefont
			{Kotliar}},\ }\bibfield  {title} {\enquote {\bibinfo {title} {Mott transition
				in the d=\ensuremath{\infty} hubbard model at zero temperature},}\ }\href
	{\doibase 10.1103/PhysRevLett.70.1666} {\bibfield  {journal} {\bibinfo
			{journal} {Phys. Rev. Lett.}\ }\textbf {\bibinfo {volume} {70}},\ \bibinfo
		{pages} {1666--1669} (\bibinfo {year} {1993})}\BibitemShut {NoStop}%
	\bibitem [{\citenamefont {Haule}(2007)}]{Haule_2007}%
	\BibitemOpen
	\bibfield  {author} {\bibinfo {author} {\bibfnamefont {Kristjan}\
			\bibnamefont {Haule}},\ }\bibfield  {title} {\enquote {\bibinfo {title}
			{Quantum monte carlo impurity solver for cluster dynamical mean-field theory
				and electronic structure calculations with adjustable cluster base},}\ }\href
	{\doibase 10.1103/PhysRevB.75.155113} {\bibfield  {journal} {\bibinfo
			{journal} {Phys. Rev. B}\ }\textbf {\bibinfo {volume} {75}},\ \bibinfo
		{pages} {155113} (\bibinfo {year} {2007})}\BibitemShut {NoStop}%
	\bibitem [{\citenamefont {Gull}\ \emph {et~al.}(2011)\citenamefont {Gull},
		\citenamefont {Millis}, \citenamefont {Lichtenstein}, \citenamefont
		{Rubtsov}, \citenamefont {Troyer},\ and\ \citenamefont {Werner}}]{Gull_2011}%
	\BibitemOpen
	\bibfield  {author} {\bibinfo {author} {\bibfnamefont {Emanuel}\ \bibnamefont
			{Gull}}, \bibinfo {author} {\bibfnamefont {Andrew~J.}\ \bibnamefont
			{Millis}}, \bibinfo {author} {\bibfnamefont {Alexander~I.}\ \bibnamefont
			{Lichtenstein}}, \bibinfo {author} {\bibfnamefont {Alexey~N.}\ \bibnamefont
			{Rubtsov}}, \bibinfo {author} {\bibfnamefont {Matthias}\ \bibnamefont
			{Troyer}}, \ and\ \bibinfo {author} {\bibfnamefont {Philipp}\ \bibnamefont
			{Werner}},\ }\bibfield  {title} {\enquote {\bibinfo {title} {Continuous-time
				monte carlo methods for quantum impurity models},}\ }\href {\doibase
		10.1103/RevModPhys.83.349} {\bibfield  {journal} {\bibinfo  {journal} {Rev.
				Mod. Phys.}\ }\textbf {\bibinfo {volume} {83}},\ \bibinfo {pages} {349--404}
		(\bibinfo {year} {2011})}\BibitemShut {NoStop}%
	\bibitem [{\citenamefont {Arsenault}\ \emph {et~al.}(2012)\citenamefont
		{Arsenault}, \citenamefont {S\'emon},\ and\ \citenamefont
		{Tremblay}}]{Arsenault:2012}%
	\BibitemOpen
	\bibfield  {author} {\bibinfo {author} {\bibfnamefont {Louis-François}\
			\bibnamefont {Arsenault}}, \bibinfo {author} {\bibfnamefont {Patrick}\
			\bibnamefont {S\'emon}}, \ and\ \bibinfo {author} {\bibfnamefont {A.-M.~S.}\
			\bibnamefont {Tremblay}},\ }\bibfield  {title} {\enquote {\bibinfo {title}
			{Benchmark of a modified iterated perturbation theory approach on the fcc
				lattice at strong coupling},}\ }\href {\doibase 10.1103/PhysRevB.86.085133}
	{\bibfield  {journal} {\bibinfo  {journal} {Phys. Rev. B}\ }\textbf {\bibinfo
			{volume} {86}},\ \bibinfo {pages} {085133} (\bibinfo {year}
		{2012})}\BibitemShut {NoStop}%
	\bibitem [{\citenamefont {Koch}\ \emph {et~al.}(2008)\citenamefont {Koch},
		\citenamefont {Sangiovanni},\ and\ \citenamefont {Gunnarsson}}]{Koch:2008}%
	\BibitemOpen
	\bibfield  {author} {\bibinfo {author} {\bibfnamefont {Erik}\ \bibnamefont
			{Koch}}, \bibinfo {author} {\bibfnamefont {Giorgio}\ \bibnamefont
			{Sangiovanni}}, \ and\ \bibinfo {author} {\bibfnamefont {Olle}\ \bibnamefont
			{Gunnarsson}},\ }\bibfield  {title} {\enquote {\bibinfo {title} {Sum rules
				and bath parametrization for quantum cluster theories},}\ }\href {\doibase
		10.1103/PhysRevB.78.115102} {\bibfield  {journal} {\bibinfo  {journal}
			{Physical Review B (Condensed Matter and Materials Physics)}\ }\textbf
		{\bibinfo {volume} {78}},\ \bibinfo {eid} {115102} (\bibinfo {year}
		{2008})}\BibitemShut {NoStop}%
	\bibitem [{\citenamefont {Bulla}\ \emph {et~al.}(2008)\citenamefont {Bulla},
		\citenamefont {Costi},\ and\ \citenamefont {Pruschke}}]{Bulla:2008}%
	\BibitemOpen
	\bibfield  {author} {\bibinfo {author} {\bibfnamefont {Ralf}\ \bibnamefont
			{Bulla}}, \bibinfo {author} {\bibfnamefont {Theo~A.}\ \bibnamefont {Costi}},
		\ and\ \bibinfo {author} {\bibfnamefont {Thomas}\ \bibnamefont {Pruschke}},\
	}\bibfield  {title} {\enquote {\bibinfo {title} {Numerical renormalization
				group method for quantum impurity systems},}\ }\href {\doibase
		10.1103/RevModPhys.80.395} {\bibfield  {journal} {\bibinfo  {journal} {Rev.
				Mod. Phys.}\ }\textbf {\bibinfo {volume} {80}},\ \bibinfo {pages} {395--450}
		(\bibinfo {year} {2008})}\BibitemShut {NoStop}%
	\bibitem [{\citenamefont {Baker}(1975)}]{Baker:1975}%
	\BibitemOpen
	\bibfield  {author} {\bibinfo {author} {\bibfnamefont {G.A.}\ \bibnamefont
			{Baker}},\ }\href
	{https://www.elsevier.com/books/essentials-of-pade-approximants/baker/978-0-12-074855-6}
	{\emph {\bibinfo {title} {Essentials of Pad{\'e} approximants}}}\ (\bibinfo
	{publisher} {Academic Press},\ \bibinfo {year} {1975})\BibitemShut {NoStop}%
	\bibitem [{\citenamefont {Vidberg}\ and\ \citenamefont
		{Serene}(1977)}]{Vidberg_1977}%
	\BibitemOpen
	\bibfield  {author} {\bibinfo {author} {\bibfnamefont {H.~J.}\ \bibnamefont
			{Vidberg}}\ and\ \bibinfo {author} {\bibfnamefont {J.~W.}\ \bibnamefont
			{Serene}},\ }\bibfield  {title} {\enquote {\bibinfo {title} {Solving the
				eliashberg equations by means ofn-point pad{\'e} approximants},}\ }\href
	{\doibase 10.1007/BF00655090} {\bibfield  {journal} {\bibinfo  {journal}
			{Journal of Low Temperature Physics}\ }\textbf {\bibinfo {volume} {29}},\
		\bibinfo {pages} {179--192} (\bibinfo {year} {1977})}\BibitemShut {NoStop}%
	\bibitem [{\citenamefont {Kohn}(1961)}]{Kohn_1961}%
	\BibitemOpen
	\bibfield  {author} {\bibinfo {author} {\bibfnamefont {Walter}\ \bibnamefont
			{Kohn}},\ }\bibfield  {title} {\enquote {\bibinfo {title} {Cyclotron
				resonance and de haas-van alphen oscillations of an interacting electron
				gas},}\ }\href {\doibase 10.1103/PhysRev.123.1242} {\bibfield  {journal}
		{\bibinfo  {journal} {Phys. Rev.}\ }\textbf {\bibinfo {volume} {123}},\
		\bibinfo {pages} {1242--1244} (\bibinfo {year} {1961})}\BibitemShut {NoStop}%
	\bibitem [{\citenamefont {Kohmoto}(1989)}]{Kohmoto_1989}%
	\BibitemOpen
	\bibfield  {author} {\bibinfo {author} {\bibfnamefont {Mahito}\ \bibnamefont
			{Kohmoto}},\ }\bibfield  {title} {\enquote {\bibinfo {title} {Zero modes and
				the quantized hall conductance of the two-dimensional lattice in a magnetic
				field},}\ }\href {\doibase 10.1103/PhysRevB.39.11943} {\bibfield  {journal}
		{\bibinfo  {journal} {Phys. Rev. B}\ }\textbf {\bibinfo {volume} {39}},\
		\bibinfo {pages} {11943--11949} (\bibinfo {year} {1989})}\BibitemShut
	{NoStop}%
	\bibitem [{\citenamefont {Naumis}(2016)}]{Naumis_2016}%
	\BibitemOpen
	\bibfield  {author} {\bibinfo {author} {\bibfnamefont {Gerardo~G.}\
			\bibnamefont {Naumis}},\ }\bibfield  {title} {\enquote {\bibinfo {title}
			{Topological map of the hofstadter butterfly: Fine structure of chern numbers
				and van hove singularities},}\ }\href {\doibase
		http://dx.doi.org/10.1016/j.physleta.2016.03.022} {\bibfield  {journal}
		{\bibinfo  {journal} {Physics Letters A}\ }\textbf {\bibinfo {volume}
			{380}},\ \bibinfo {pages} {1772 -- 1780} (\bibinfo {year}
		{2016})}\BibitemShut {NoStop}%
\end{thebibliography}
%\bibliographystyle{Style_bibliography}

%merlin.mbs apsrev4-1.bst 2010-07-25 4.21a (PWD, AO, DPC) hacked
%Control: key (0)
%Control: author (0) dotless jnrlst
%Control: editor formatted (1) identically to author
%Control: production of article title (0) allowed
%Control: page (1) range
%Control: year (0) verbatim
%Control: production of eprint (0) enabled
%

\end{document}